\documentclass[12pt]{article}
\usepackage{amsmath,amssymb,graphicx,mathrsfs,hyperref,slashed,geometry}
\newcommand{\be}{\begin{equation}}
\newcommand{\ee}{\end{equation}}
\newcommand{\bea}{\begin{eqnarray}}
\newcommand{\eea}{\end{eqnarray}}

\def\({\left(} \def\){\right)}

\renewcommand{\baselinestretch}{1.25}
\begin{document}
\newgeometry{left=1.5cm,right=1.5cm}
\title{\vspace{-1.8in}
{Non-singular black holes interiors  need \\ physics beyond the standard model}}
\author{\large Ram Brustein${}^{(1,2)}$, A.J.M. Medved${}^{(3,4)}$
\\
\vspace{-.5in} \hspace{-.05in}  \vbox{
\begin{flushleft}
$^{\textrm{\normalsize
(1)\ Department of Physics, Ben-Gurion University,
Beer-Sheva 84105, Israel}}$
$^{\textrm{\normalsize
(2)\ Theoretical Physics Department, CERN, 1211 Geneva 23, Switzerland}}$
$^{\textrm{\normalsize (3)\ Department of Physics \& Electronics, Rhodes University,
Grahamstown 6140, South Africa}}$
$^{\textrm{\normalsize (4)\ National Institute for Theoretical Physics (NITheP), Western Cape 7602,
South Africa}}$
\\ \small \hspace{1.07in}
ramyb@bgu.ac.il,\ j.medved@ru.ac.za
\end{flushleft}
}}
\date{}
\maketitle
\begin{abstract}
The composition as well as the very existence of the interior of a Schwarzschild  black hole (BH)  remains at the forefront of interesting, open problems in fundamental physics. To address this issue, we turn to  Hawking's ``principle of ignorance'', which says that, for an observer with limited information about a physical system, all descriptions that are consistent with known physics are equally valid. We compare three different observers who view the BH from the outside and agree on the external  Schwarzschild geometry. First, the modernist, who accepts the classical BH as the final state of gravitational collapse, the singularity theorems that underlie this premise and  the central singularity that the theorems predict. The modernist is willing to describe matter in terms of quantum fields in curved space but insists on (semi)classical gravity. Second is the  skeptic, who wishes to evade any singular behavior by finding a loophole to the singularity theorems within the realm of classical general relativity (GR). The third is a postmodernist who similarly wants to circumvent the singularity theorems  but is willing to invoke exotic quantum physics in the gravitational and/or matter sector to do so. The postmodern view suggests that the uncertainty principle can stabilize a classically singular BH in  a similar manner to the stabilization of the classically unstable hydrogen atom: Strong quantum effects in  the matter and gravitational  sectors   resolve the  would-be singularity over horizon-sized length  scales. The postmodern picture then requires a  significant departure from (semi)classical gravity, as well  as some exotic matter beyond the  standard model of particle physics (SM). We find  that only the postmodern framework is consistent with what is known so far about BH physics and conclude that a valid description of the BH interior needs matter beyond the SM and  gravitational physics beyond (semi)classical GR.
\end{abstract}
\restoregeometry
\renewcommand{\baselinestretch}{1.5}\normalsize


\subsection*{Introduction}

What is the final state of matter that collapses under its own gravity?  The answer  has long  been  debated, as far back as  the early days of Newtonian gravity.  When general relativity (GR) came along, the debate raged on, but now in terms of  the framework of the new theory. Finkelstein eventually realized that the Schwarzschild solution was describing an event horizon \cite{fink} and was later  revealed to be truly singular.  Many  attempted to prove otherwise and show that a  singularity cannot be part of the correct description of the final state of collapsing matter.  A period of intense debate then ensued between two opposing camps: Those unwilling to accept the demise of physics by its own hands, such as Wheeler \cite{geons},  versus  the likes of Buchdahl \cite{Buchdahl}, Chandrasekhar \cite{chand1,chand2} and  Bondi \cite{bondi}, who applied the equations of  GR  to show that ``normal" matter cannot be stable when confined to some minimal radius. In parallel, more formal efforts at proving the inevitability of singularities were pursued by Raychaudhuri \cite{ray}, Komar \cite{komar} and others. The debate ended (or so it seemed) when Penrose and Hawking  proved their singularity theorems  \cite{PenHawk1,PenHawk2}. Soon thereafter, the term ``black hole" was invented  and some semiclassical aspects  were added by Bekenstein \cite{Bek} and Hawking \cite{Hawk,info}; thus establishing the paradigm of
 black hole (BH) thermodynamics and marking the beginning of the modern era of BH physics.

From the modern perspective, an observer should see nothing unusual as she falls through a BH horizon; after all, this is a region of weak gravity for a BH that is sufficiently  massive.~\footnote{A large Schwarzschild BH in a four-dimensional, asymptotically flat spacetime is assumed throughout.}  This seemingly uncontroversial opinion is dependent on at least two implicit assumptions; namely that our normal understanding of spacetime geometry persists into the BH interior and there is a  separation of scales between the central singularity and
the horizon at the  gravitational radius.  Although both seem reasonable enough contentions, there are no grounds for insisting that they {\em must be}  true.  The BH interior is by definition causally separated from the rest of the Universe, and there are reasons to suspect that the modern perspective is amiss, as it invokes  a central singularity, as well as a reversal of time and space.

If there is no clear-cut picture of the interior, then what is one to do? Hawking implicitly answered this question long ago with a  brief discussion on what he called  the  ``principle of ignorance'' \cite{info}. To paraphrase, an observer who is lacking information about a physical situation is free to adopt any explanation that is consistent with the laws of physics and experimental data, as all such  descriptions are equally likely.  With the adoption of this principle, the  portrait of a BH in GR is as valid as one's tolerance of singular regions of spacetime.
And,  as discussed below, a regularized singularity, the frequently  prescribed remedy, is just as flawed as a mathematical infinity.

The task of constructing a viable model for the interior becomes all the more challenging when quantum effects are incorporated; in particular, the quantum process of BH evaporation \cite{Hawk}. It  has  long been understood that   an evaporating BH --- for which the final outcome appears to be  a maximally mixed state of thermal radiation --- is  inconsistent with the  quantum principle of  unitary evolution \cite{info}. There was a time when the preferred way out of this dilemma relied upon a generalized  notion of Bohr-like complementarity \cite{comp1,comp2,comp3}, which claimed that causally separated observers need not agree on physical events as long as they can never compare their observations. And so, on this basis,   the same information  can simultaneously be both inside and outside of a BH simply because no single observer could verify this duplicity.
(If she could, this would contradict a quantum theorem against  cloning.) Our own current view (see below) is that the interior observer  gets to have  no say   and that this is how observer complementarity can be set aside.

Nevertheless, it has since been made clear that, given a unitary evaporation process,  a single external observer can still see a violation of yet another cherished principle of quantum mechanics: the monogamy of entanglement or, as  it is more formally known, the strong subadditivity of the  von Neumann entropy. This conflict  has been  brought to  light in \cite{Sunny,Mathur1,Braun,MP,Mathurtalk} and, most famously, by the authors of \cite{AMPS} (AMPS). They  ``doubled down'' on the GR model  by moving the  BH singularity all the way out to  where the horizon should be, thus providing the BH with a metaphorical ``firewall''. To be fair,  AMPS   were  not as much  advocating  for the existence of a firewall as they were illustrating the fallacy of using observer  complementarity (at least in this context).

The  advent of the firewall argument  marked the beginning of the postmodern era of
BH physics. The essence of the firewall argument is that it is not reasonable to simply  ignore the interior and the ensuing challenge for the postmodern era is, succinctly, the singularity.

There has since  been a long line of attempts at circumventing the AMPS   solution  while, at the same time,  having one's unitary evolution and strong subadditivity too. (See \cite{AMPSS} for an already exhausting list of what is but a small fraction of the ensuing papers.) More often than not, the singularity is  treated as  a triviality under the premise (sometimes implicit) that quantum gravity or string theory should somehow regularize any infinities  and thus  save the day. An additional implicit assumption is that the regularized singularity will not affect the structure of space time on horizon-sized scales. The problem with this mindset is that  a regularized singularity is no more or less than a BH remnant, as both imply an extremely large amount of  stored information is a small region of space. And, although arguments for remnants still turn up from time to time, it is widely accepted that their putative existence opens up the Pandora's box of a destabilized vacuum \cite{sussremn}  from which one should promptly move on. For a recent review on remnants, see \cite{remrev}.

A notable exception  is Mathur's string-theory-inspired fuzzball model \cite{MathurFB} (also, \cite{otherfuzzball}), which  considers a quantum superposition of singularity-free and horizon-free microstates. The total wavefunction is then supposed to mimic the properties of a semiclassical BH. Another exception is 't~Hooft's BH without an interior
\cite{HooftX,HooftXX}, which is based on the idea that high-energy incoming particles can be traded for low-energy outgoing particles (these being quantum  clones of one another), meaning that the interior need never be probed.
Yet another exception is the graviton-condensate model of Dvali and Gomez \cite{Dvali}, who invoke a highly-occupied state of gravitons as the state of the BH to obtain a regular interior with  only an approximate horizon.

Meanwhile,  another issue with BH evaporation has recently emerged; what Mathur has called the BH ``causality paradox'' \cite{mathursupernew}. The challenge here is to explain how information can escape out of a BH given that the future light cone of any interior particle is completely contained within the horizon. This presents an additional hoop to jump through for those with a vested interest.

From our point of view, the singularity/remnant problem, the tension between strong subadditivity and unitarity, and the apparent conflict with causality represent different  sides of the same three-sided coin. As long as there is a singularity, then information loss is inevitable because absorption by the singularity is effectively the same  as passing through to another universe  as in Wheeler's  bag-of-gold spacetime \cite{gold}. Conversely, any model that is based upon the rules of quantum theory and devoid of singularities  must, by construction,
respect unitarity,  strong subadditivity and  causality, as well as any other fundamental tenet of quantum theory. If gravitational collapse can somehow be avoided, then the rest will follow.

Our proposed resolution is that the uncertainty principle can stabilize a classically singular  BH in  essentially the same way that it  can do so for a classically unstable atom. For an early discussion of this idea see Sect. 1.3 of \cite{davies}.  We expect  strong quantum effects from  the matter and gravitational  sectors  to  ``smear" the  would-be singularity over  horizon-sized length  scales. The self-consistency of this picture of the  interior   requires a  significant departure from (semi)classical gravity, as well  as some exotic matter which is outside  the
realm of the  standard model (SM). The resulting picture is that of a ``quantum star" that looks from the outside just like a BH.

But let us circle back to the beginning of the discussion and recall that one story is as good as another as long as it consistent with what is already known about physics. In light of this, we would like  to understand the interior from the perspective of an external observer. The distinction  between interior and exterior observers is more than semantics because different perspectives could have  provided, in principle,  complementary descriptions (in the quantum sense) of the same physical system. For instance, if Alice probes a system with operators that are non-commuting with respect to Bob's, this ubiquitous pair of observers  will necessarily have conflicting descriptions. The essential  difference between BHs and most other systems is that, in the BH case, the perspective of an external observer is the only one that  really counts. This is, in our opinion, the essence of why BH complementarity as it is usually interpreted  is problematic; the interior
observer  gets to have  no say in the matter.

With all of this in mind, our objective is to describe
the BH interior from the perspective of three different outside observers, each with their own distinct narrative. These will be (1) the modernist who adheres to the modern view of a  BH, that of GR including BH thermodynamics, and is willing to accept a (regularized) singularity as part of her worldview and therefore abandon unitarity,
(2) the skeptic who  allows  exotic matter if its inclusion evades gravitational collapse, views BH thermodynamics and unitarity as desirable although non-essential  but otherwise  abides by the rules of classical GR and
(3) the  postmodernist who abhors a singular gravitational collapse, insists on unitarity  and  will call upon
exotic physics in both the matter and gravitational sector as needed to complete a paradox-free picture.

We discuss each of these in turn and then provide an overview at the end.

\subsection*{The modern perspective}

Here, we will describe the viewpoint of an external observer who insists on the BHs of GR as a starting point, while incorporating the commonly accepted ideas from the realm of semiclassical physics. This observer assumes gravitationally collapsing matter in the usual sense, except
that the endpoint is some sort of regularized  singularity or remnant. The firewall model would also fall under this category, as it at least starts out with the traditional (semiclassical) picture, and nothing unusual has to happen until  the midpoint of BH evaporation as measured in units of
diminishing entropy \cite{AMPS} (what is  known as the Page time \cite{page}).

\begin{figure}[htb]
\centerline{\includegraphics[width=10.5cm]{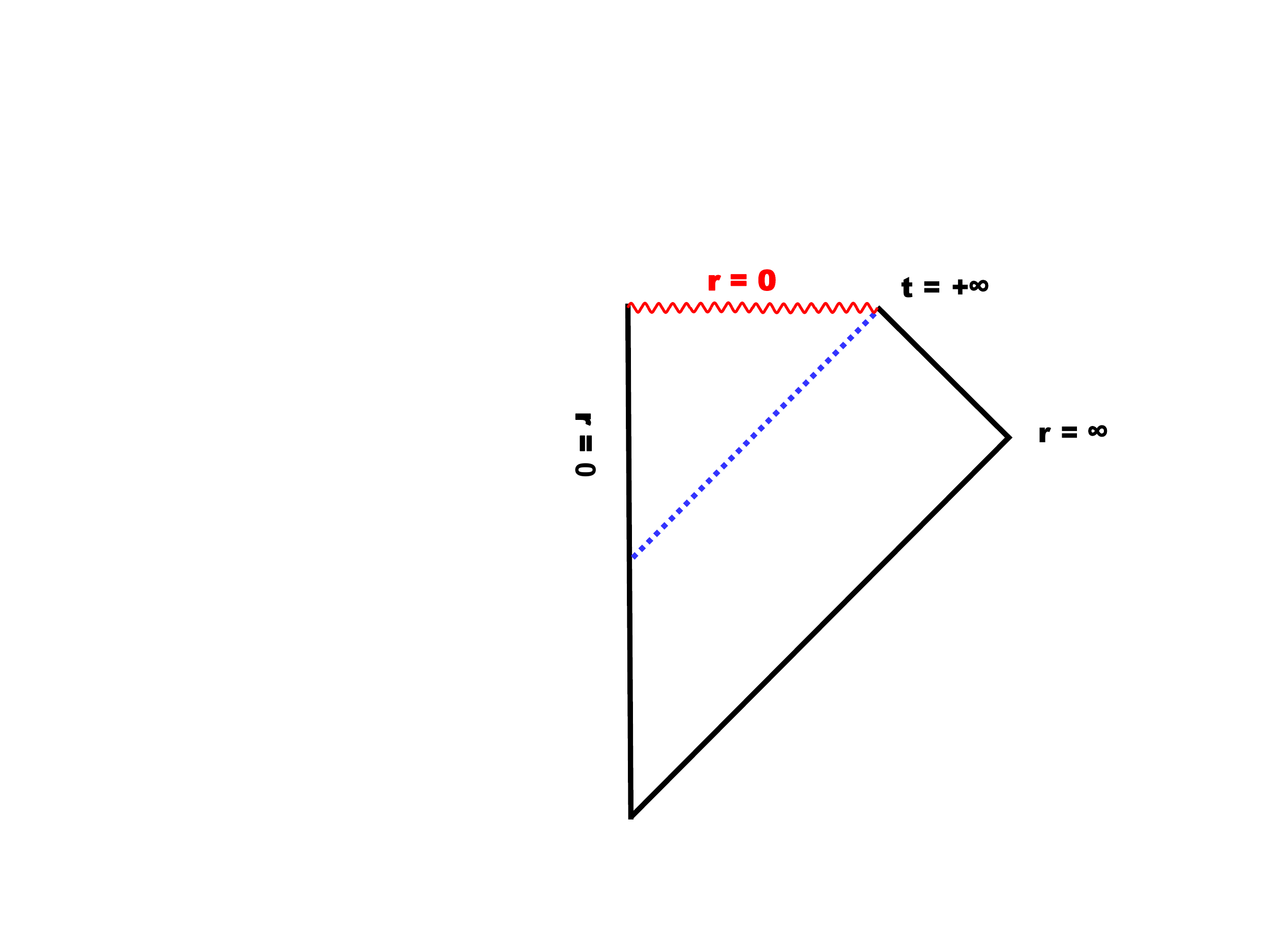}}
\caption{Penrose diagram of the modernist view of the BH, excluding evaporation.}
\label{fig:modernist}
\end{figure}

As this picture of a BH is (almost) singular by design, the observer must be
assuming  some notion  of  a   remnant  is consistent with known physics. This observer must also  accept  the  separation of scales between the singularity (or some regularized version of it)  and the horizon, so that the former scale  has no bearing on  the latter. Although such a stance is  highly questionable at best,  we will assume for the sake of discussion that these are acceptable conditions and proceed to consider other aspects of the modernist's description. A Penrose diagram depicting the geometry (excluding evaporation) from the modern perspective is depicted in Fig.~\ref{fig:modernist}.

In view  of the causality paradox,  the observer must be of the impression  that neither matter nor  information could escape from the BH interior. If the observer still insists on BH radiation, as well she should, the obvious explanation is an external pair-production process just as Hawking described it  in \cite{info}. For any given pair, the positive-energy partner transitions into an emitted Hawking particle whereas the negative-energy partner passes through the horizon and consequently lowers the mass of the BH.  There is, however,  a
wrench in the pair-producing gears: One must inevitably choose between either entangled pairs at the horizon but with an accompanying loss of information or, else, the purification of the radiation but at the prohibitive cost of a firewall, which is tantamount to moving the singularity all the way to the horizon. This choice suggests that the  separation of BH scales may not be such an valid idea after all.

One way out of this conundrum is if the negative partner's ``half'' of the entanglement  could be teleported or swapped to the external radiation; a protocol that was recently  outlined in \cite{swap}. The good news is that, just   like any other teleportation event,  there can be  no violation of causality. The not-so-good news is that, in this case, the swap is to take place just when the negative partner is to be annihilated, which would be much more tenable if it did not involve the BH singularity. And even though  there has been a proposal to this effect --- namely, the final-state solution \cite{HorMal} ---  it has been revealed to be a non-unitary process unless  there is a disturbing amount of fine-tuning \cite{nonHorMal}.

Another way out would be  via a traversable wormhole in the case of a two-sided BH or, more generally and by the same logic,  the  ER=EPR proposal \cite{MalSus}. In the latter scenario, the interior of the BH is connected to the external radiation via of an  Einstein--Rosen bridge with multiple exits; any of
 which could be far away from any of the others  in terms of  real space. This bridge maintains the near-horizon  entanglement as long as necessary  while  providing a conduit for the information to eventually flow out. It is, in fact,  not totally clear whether or not the ER=EPR picture  falls more in line with  postmodernist reasoning because, when the length of the ER bridge reaches the order of the horizon radius, the separation-of-scales assumption is no longer applicable (as would also  be true for  any postmodern  model).   On the other hand, the  original proponents of  ER=EPR suggest that it provides an alternative interpretation of the seminal   firewall model (see Sections~4.3 and~6 in \cite{MalSus}), and so we also include it here.

In any event, insofar as  the Hawking particles originate strictly  on the outside of the horizon, the associated entropy would have to   be attributed to the thermal atmosphere of the BH. As shown by 't Hooft \cite{BW}, this entropy would indeed agree with the BH area law \cite{Bek,Hawk}. Nevertheless, the actual BH entropy would have to be located in the singularity, as this is where all the collapsed matter ends up. And so  the modernist's story takes us right back to the remnant idea and its associated failings. In short, one must be willing to suspend  disbelief to conclude that this problematic description of the interior is  consistent with  the known laws of  physics.

\subsection*{The skeptic's perspective}

As already stated, this would be the viewpoint of an external observer who
proposes to avoid collapse by incorporating exotic matter but, at the same time, wants to stay within the realm of classical GR. The exotic matter need not be an actual form of matter; for instance, it could be some  classical terms in the stress tensor  that are meant to mimic corrections from quantum gravity. Regardless, because of the adherence to GR,  any such observer will have to confront the Buchdahl bound \cite{Buchdahl}.

Buchdahl considered the stability (or lack thereof) of a spherically symmetric matter distribution.
Implementing only causality and   a handful of common-sense assumptions
about the energy density $\rho$ and pressure $p$, he was able to establish that
stability required a minimal radial size  of 9/8 Schwarzschild radii. Even though our focus will be on  Buchdahl's calculation,  similar reasoning led to the same  general conclusions in some contemporary articles  and the same basic principles are the essence of  the famous singularity theorems (see the Introduction).

What was for a long time overlooked, although noticed and then summarily dismissed by Bondi \cite{bondi}, is that a maximally negative radial component of  pressure, $\;p_r=-\rho\;$,  is sufficient to bypass the Buchdahl bound, provided that the matter remains sufficiently dense  all the way up to the Schwarzschild radius
(this caveat will be clarified below).
But it was eventually realized  that negative pressure is indeed a  key ingredient  for evading the Buchdahl bound (as well as  the singularity theorems) and  has since been incorporated
into  many attempts at  modeling  the interior of  a regularized BH.   Generically, such models are based on the idea that the deviations from GR only occur at scales where quantum-gravitational corrections become important. These models describe objects for which the deviations are limited to some scale $\;R_{QG}\ll R_S\;$, where $R_S$ is the object's Schwarzschild radius.

\begin{figure}[htb]
\centerline{\includegraphics[width=10.5cm]{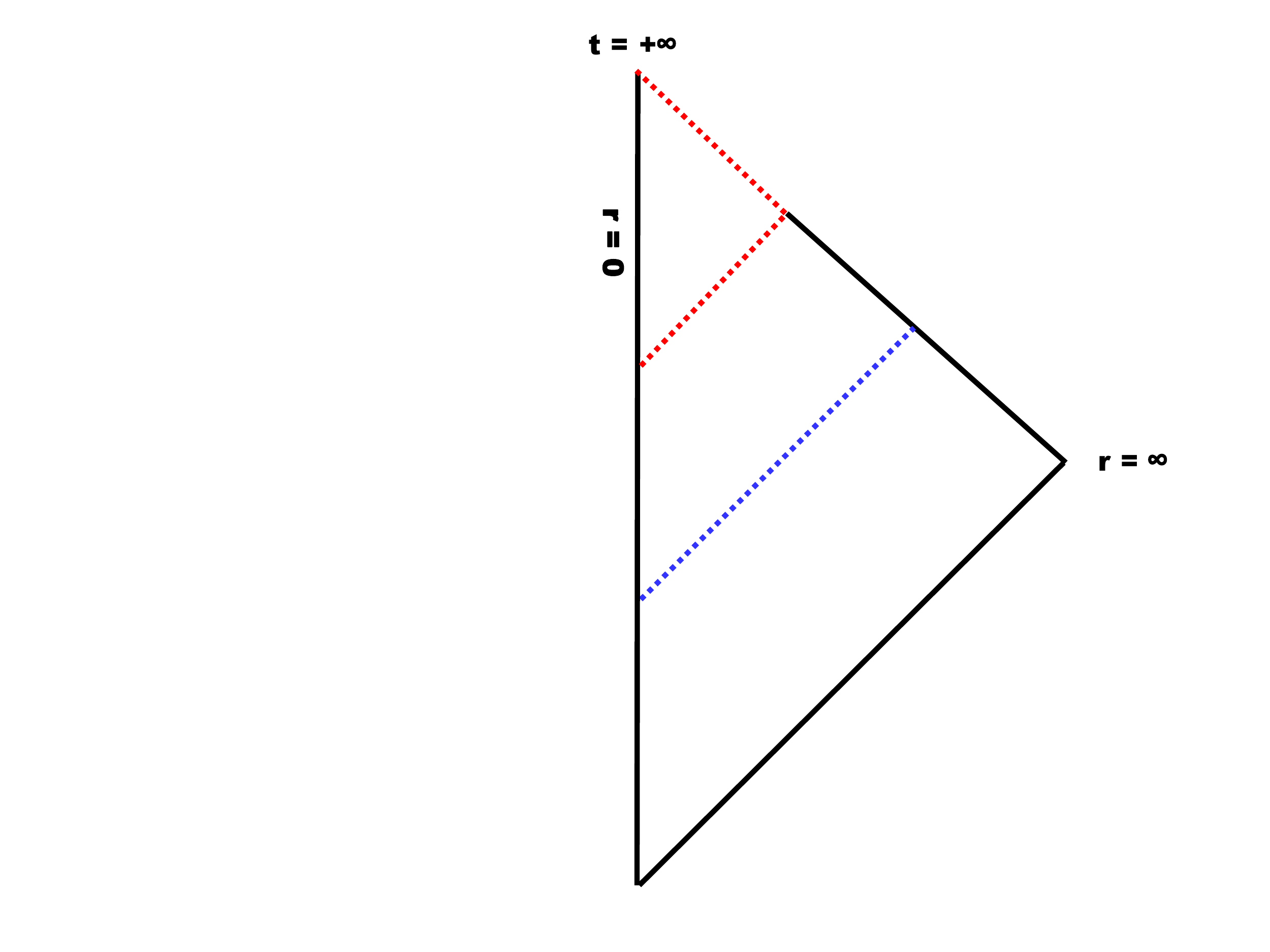}}
\caption{Penrose diagram of the Hayward model of a regular BH, representing a skeptic's perspective. }
\label{fig:hayward}
\end{figure}

A prototype model of this kind is that of Hayward \cite{hayward}. The geometry in this model is defined via the line element
\be
ds^2\;=\; -f(r) dt^2 + \frac{1}{f(r)} dr^2 + r^2 d\Omega_2^2 \;,
\ee
for which
\be
f(r)\;=\;1- \frac{2 mr^2}{r^3 + 2 m L^2}\;,
\ee
and the corresponding  Penrose diagram is shown in Fig.~\ref{fig:hayward}.
The length scales $m$ and $L$ determine the location of a pair of a horizons; an outer one at  $\;r\sim 2 m\;$ and an inner horizon at $\;r\sim L\;$ (the latter also plays the role of $R_{QG}$).
The Hayward model features a negative radial pressure  $\;p_r=-\rho\;$ and a
transverse pressure $p_{\perp}$ that is  negative near the center but  positive
for $r>L$. The energy density and pressure are both parametrically small outside of  the inner horizon.

\begin{figure}[htb]
\centerline{\includegraphics[width=10.5cm]{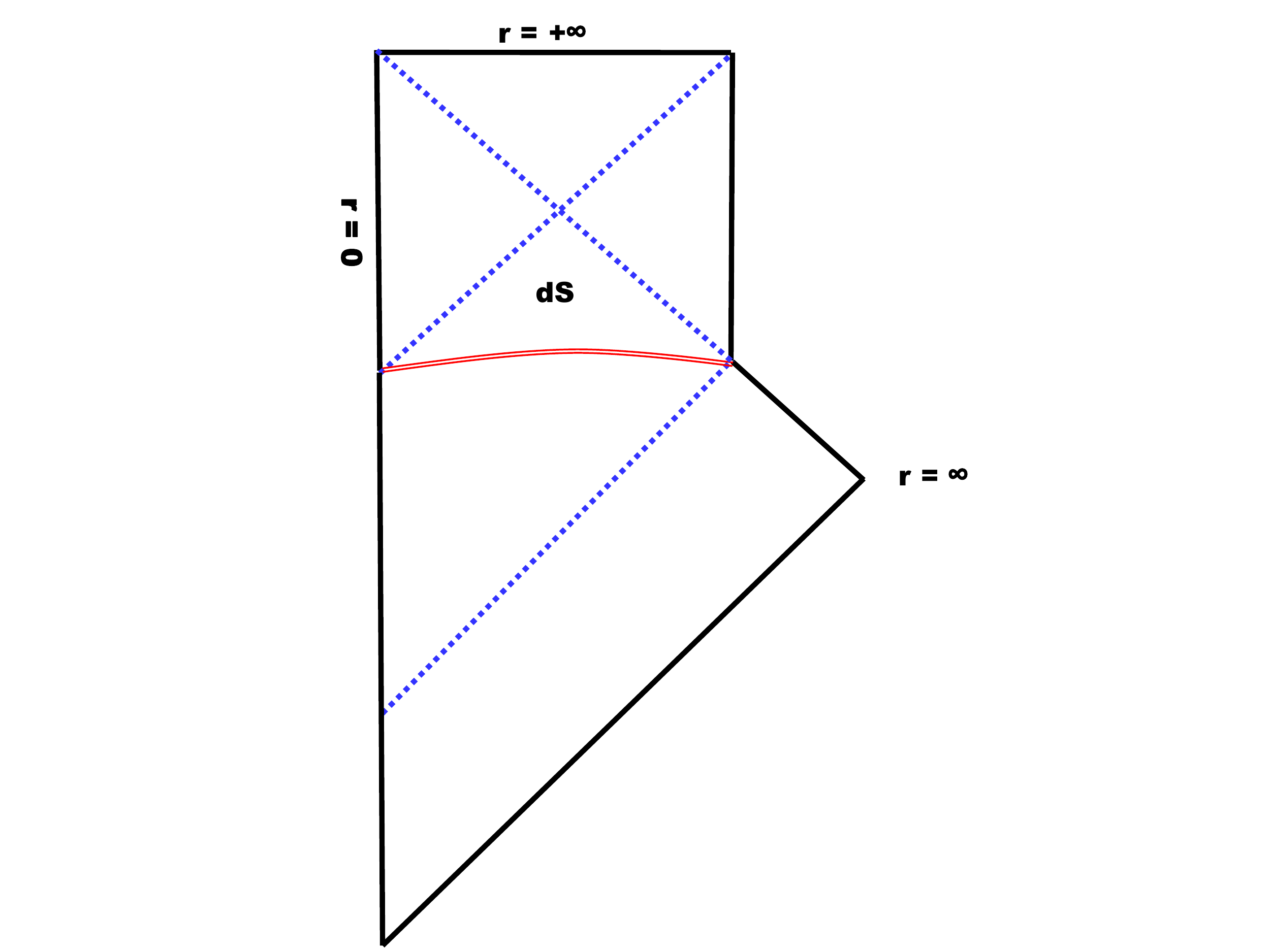}}
\vspace{-.1cm}
\caption{Penrose diagram of the FMM model of a regular BH, representing another  perspective of a skeptic.}
\label{fig:fmm}
\end{figure}

The Frolov--Markov--Mukhanov (FMM) model employs the same basic idea as
Hayward's, but with the  inner horizon  replaced by a shell which connects a truncated Schwarzschild interior to an entire de Sitter universe in a smooth way \cite{FMM}. The Penrose diagram corresponding to this geometry is illustrated in Fig.~\ref{fig:fmm}.  Other  models of this ilk ---
 exploiting the idea of negative pressure for the purpose of regularizing the interior ---
 are the gravastar \cite{MMfirst}, the  black star  \cite{barcelo}, a back-reaction-corrected geometry  \cite{CR} and  our own model, the collapsed polymer BH \cite{bookdill}, as well as  \cite{eg1,eg2,eg3}. See \cite{frolovrev,splucc,eg4} for extensive reviews of models of regular BHs.  Apparently, the earliest example of a non-singular BH can be attributed to Bardeen \cite{bardeen}, who  rather based his model on a
core of charged matter. A recent  version of Bardeen's model can be found in  \cite{AyonBeato:1998ub}, which has the same causal structure as that of the Hayward model.

But there is a major issue with most of these regularized BH models
that comes to the fore when the evaporation process is taken into account; the earliest examples of which can be found in \cite{FV,hayward}. As explained in  \cite{frolov,visser}, it is a generic feature of regular BH solutions
with $\;R_{QG}\ll R_S\;$ that the energy of their  emitted particles can sum up to an  amount which is much larger than the object's original mass. Of course, if one is willing to ignore quantum mechanics completely, the problem goes away.  We do, however, view this as an indication that such  models are fundamentally inconsistent.

What is then needed  is an object whose interior metric never approaches the Schwarzschild solution until it is parametrically close to the horizon; that is, $\;R_{QG}\sim R_S\;$.  From this perspective, our own  polymer model and the gravastar  stand out because, in these two cases, the deviations persist up to a radius that is parametrically close to  the outer surface
\cite{MMfirst,bookdill}. We will then turn our attention to this pair of
models, beginning with the gravastar.

\begin{figure}[htb]
\centerline{\includegraphics[width=10.5cm]{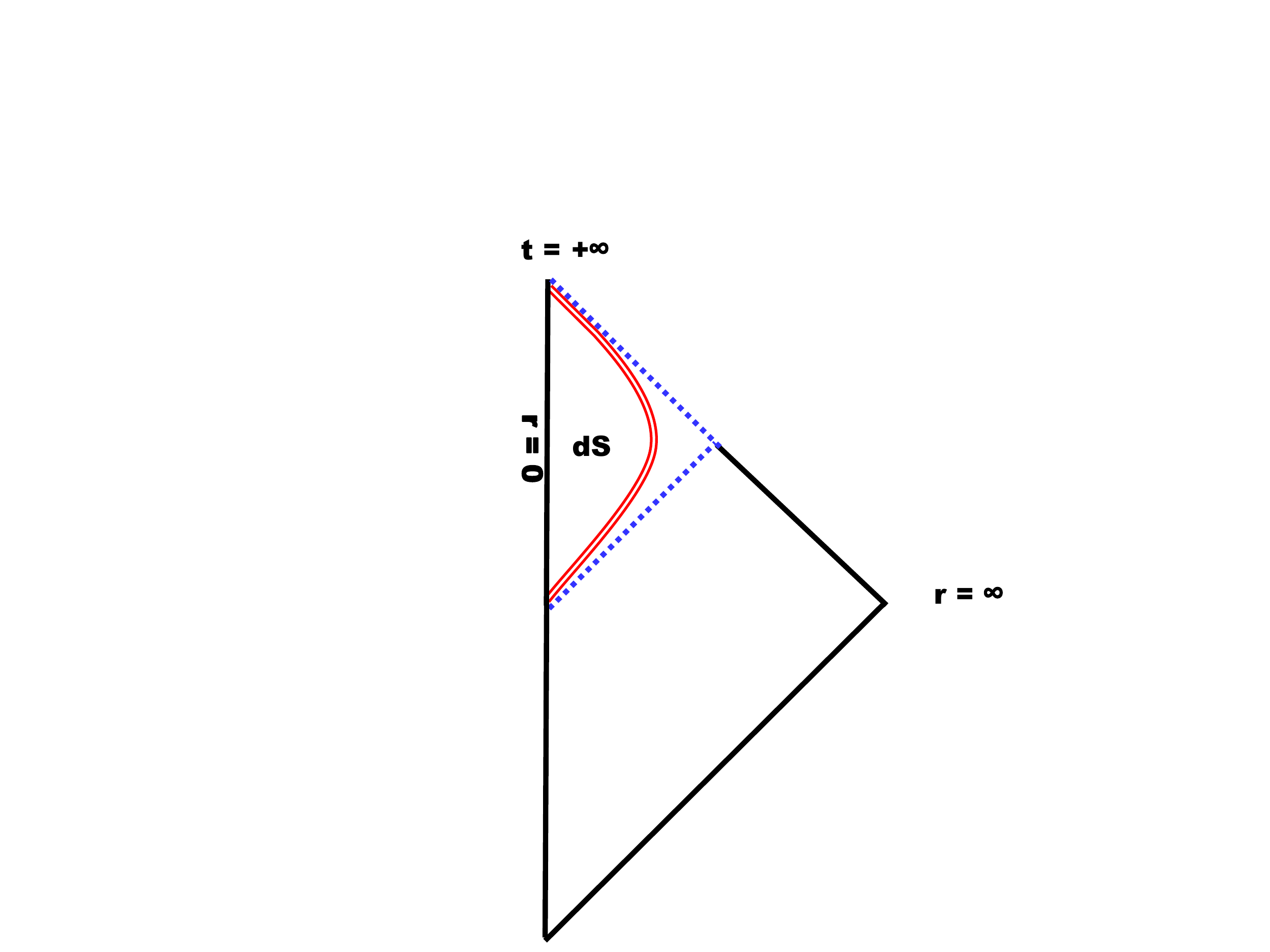}}
\vspace{-0.05cm}
\caption{Penrose diagram of the gravastar model of a regular BH, representing yet another skeptic perspective. }
\label{fig:gravastar}
\end{figure}

 For Mazur and Mottola's gravastar model \cite{MMfirst}, the interior has a constant energy density and a constant, isotropic pressure \cite{MM}. This solution describes an ultracompact object with what is  essentially a de Sitter interior along with an  outer  shell of matter. The latter is necessary to ensure that the de Sitter interior can be matched smoothly to the  exterior Schwarzschild solution. The interior geometry of this model is described by the  line element
\be
ds_{int}^2 \;=\; -\frac{1}{4} \left(1- \frac{r^2}{R_S^2}\right) dt^2 + \frac{1}{1- \frac{r^2}{R_S^2}} dr^2 + r^2 d\Omega_2^2\;,
\ee
and the Penrose diagram  for the gravastar is presented  in Fig.~\ref{fig:gravastar}.

Our own proposal --- the collapsed-polymer model \cite{strungout} ---  similarly has a maximally negative pressure, but only  from the perspective of a skeptic. This subtle point regarding observer dependence was touched  upon in \cite{bookdill}
and will be elaborated on later when we discuss the same model from a postmodern perspective.
Meanwhile, as far as the  skeptic is concerned,
\be
\;p_r \;=\; - \rho \;=\; - \frac{1}{8\pi G r^2}\;
\label{pr}
\ee
along with a vanishing transverse pressure \cite{bookdill}.~\footnote{In spite of appearances, this is a non-singular matter distribution as
the physically relevant quantity, $dr\;4\pi r^2 \rho$, is finite throughout.}  A model that features  $\;p_r = - \rho\propto - \frac{1}{r^2}\;$ was discussed in another context \cite{guendel1,guendel2} in which the negative pressure was sourced by a spherically symmetric ``hedgehog" configuration of cosmic strings.

\begin{figure}[htb]
\centerline{\hspace{-6cm}\includegraphics[width=13.5cm]{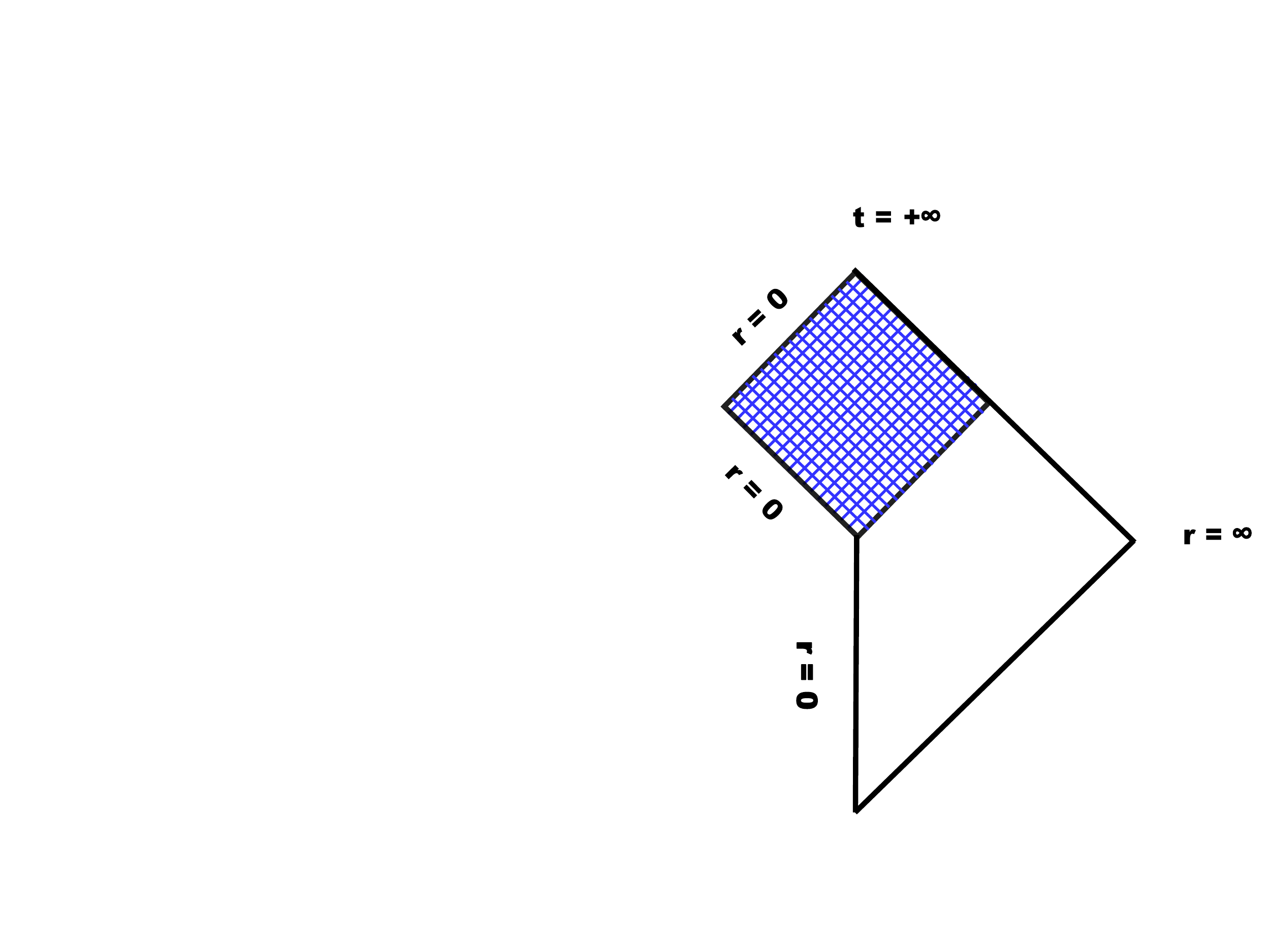}}
\caption{Penrose diagram of the polymer model of a regular BH from a skeptic's  perspective.
It should be viewed, as explained in \cite{auntiewho}, as part of an extended diagram.
The whole interior is null, each sphere is like a BH horizon. See Fig.~\ref{fig:polymerPM} for the  postmodern perspective of the same model.}
\label{fig:polymerskeptic}
\end{figure}

The interior line element of the  skeptic's version of the polymer  geometry is given by
\be
ds^2\; =\;- f(r) dt^2 +\frac{1}{f(r)} dr^2 +r^2 d\Omega_2^2\;,
\ee
such that  $\;f(r)=0\;$. The metric then adopts the peculiar form of
$\;g_{tt}=g^{rr}=0\;$ throughout the interior; implying that the whole region
is null.  Every spherical surface of constant radius is itself just like a BH horizon!  The Penrose diagram corresponding to this model is depicted in Fig.~\ref{fig:polymerskeptic}.

A remarkable aspect of the skeptic's polymer  model is its  complete stability against quantum perturbations \cite{bookdill}; meaning that there is no opportunity for the spontaneous emission of Hawking radiation. This is consistent with having zero entropy, as a classical solution should. This is fine with the skeptic since his
preference would be  to rely on classical gravitational physics as much as possible,  even if certain terms in the Lagrangian are meant to imitate the effects of quantum gravity.

Nevertheless, an external
perturbation that violates the null energy condition (as is perfectly reasonable in a quantum theory~\footnote{A quantum description of an  exterior matter system is acceptable to the skeptic.}) will produce radiation, albeit in the form of gravitational waves, and also some modified Hawking radiation
\cite{newwithYoav,YotamNew}. And, because of this violation, faster-than-light travel can be anticipated, as discussed recently in \cite{{Itzhaki:2018glf}}.
The causality paradox is then evaded in this case  in a way that does not contradict the SM or quantum physics.
In fact, as explained in the next section, the external perturbation does not need to violate any energy condition at all for the purposes of bypassing the causality bound.
Nevertheless, according  to a skeptic, an external source is indeed necessary for the polymer model because of the otherwise stable interior.

The way to understand the previous claim of faster-than-light travel is to notice that light-like travel of incoming/outgoing perturbations can be expected in a medium for which $\;p_r^2=\rho^2\;$ holds identically because then the radial speed $v_r$ is unity,
$\;\frac{v^2_r}{c^2}= \left|\frac{dp_r}{d\rho}\right|=1\;$, and that the null version of the Raychaudhuri equation (along with Einstein's equations) tells us that light rays will defocus  only if the null energy condition is violated. This becomes an ``if and only if'' statement provided that  the associated shear, torsion and expansion $\Theta$  are all small enough to neglect because, in this
case,
\be
\frac{d\Theta}{d\lambda} \;=\; -\left(\rho + p_r\right)\;>\; 0\;,
\ee
where $\lambda$ is the relevant  affine parameter.

Irrespective of the causality paradox, this perspective of the interior still  presents a significant problem: the conspicuous absence of an entropy, at least one that is large enough to account for the BH area law. This state of affairs applies just as well to  the gravastar as it does to the skeptic's version of
the polymer BH, and quite possibly to any other self-consistent model with a
classical and regularized interior. To resolve this issue,  one could resort to treating the entropy as a mechanical quantity, in  much the same way that the Wald entropy is introduced \cite{WaldEntropy}.  This is, in fact, the approach
that was adopted for  the  gravastar model  in \cite{MM}.  Meanwhile,  the vanishing of the thermodynamic entropy is necessarily true given that $\;Ts=\rho+p=0\;$ as it  is for the gravastar  or, in the case of the skeptic's polymer model,
 $\;Ts=\rho+p_r=0\;$,~\footnote{The reason that $p_r$ is used here is that BH horizons effectively act like $(1+1)$-dimensional surfaces and every spherical surface in this interior acts like a horizon. Alternatively, the postmodern  version of the polymer BH is filled with closed strings \cite{strungout}, and so any excitation ``sees'' a $(1+1)$-dimensional geometry.} where $s$ is the entropy density and $T$ is the temperature. The alternative  of a vanishing temperature would be  equally unwelcome. And so, even though there may be an emission of radiation, its temperature would not have to agree with the standard paradigm of BH thermodynamics.

In summary, the skeptic's  story is fine at the level of classical gravitational physics but is ill suited for the incorporation of quantum effects. Indeed, the various models of regularized BH solutions are fine as long as evaporation and  other related quantum aspects are not considered, as  previously discussed. However, when one attempts to include quantum effects, inconsistencies abound.

\subsection*{The postmodern perspective}

Just like a skeptic, a postmodernist is determined to eliminate  the singularity from her portrait of the BH interior. The difference here is that the postmodernist is willing  to extend the bounds of conventional physics; for instance, quantum physics extending gravity beyond classical GR, exotic matter extending field theory beyond the SM  or likely  both. The prototypical example of the postmodern perspective is the string-theory inspired fuzzball model \cite{MathurFB} .
Other prominent examples are 't~Hooft's BHs without interiors \cite{HooftX,HooftXX}, the graviton-condensate model of Dvali and Gomez \cite{Dvali} and, depending on its classification, Maldacena and Susskind's ER=EPR proposal. Except for a brief discussion at the end of the section, much of  our focus will again be on the polymer model of the BH, although now from the postmodernist's  perspective. The polymer model, just like the others, relies on properties of of string theory and describes deviations over horizon-sized scales, but it also provides  a direct link to the discussion in the previous section.  Another example of the link between a skeptic and a postmodernist perspective is described in \cite{GermaniCesc}. There, the graviton condensate model is mapped on the gravastar model.

The polymer model grew out of the notion that the BH interior must be in a non-classical state, even at times before the Page time. This claim has been a  common theme in some of our recent work, beginning with \cite{density},  but the most explicit argument is presented in \cite{noclass}. There, it is shown that the BH radiation is in a highly quantum state when expressed in terms of the Fock (or occupation number) basis of asymptotic quantum fields. It follows  that the purifier of the radiation --- the BH interior --- is of a similarly quantum nature when expressed in the corresponding basis (this is also shown explicitly in \cite{noclass}). Our conclusion is that a  geometrical description of the interior in terms of a semiclassical metric is not feasible. Simply put, the quantum fluctuations in the metric would be as least as large as the corresponding  expectation values.

The premise of the polymer model is  that the BH  interior consists of  highly excited, long, closed, interacting  strings \cite{inny,strungout}. The
primary motivation  is that the natural distribution of matter for a
highly  quantum interior happens to have the same equation of state  as a collection of long, closed strings when heated to just above the Hagedorn temperature, as included below for completeness. What closes this circle of logic is   that such strings have a high density of states and, as such,  are  subject to exceptionally large quantum fluctuations \cite{Deo}.

The equation of state for this high-temperature string phase is famously $\;p=\rho\;$ \cite{AW}. In such a phase, the entropy density $s$
and temperature $T$ are related  to the energy density, in string units, as  $\;s=\sqrt{\rho}\;$ and  $\;1/T=ds/d\rho=s/2\rho\;$. Then $\;p=-\rho+sT= +\rho\;$ which is, of course, consistent with the thermodynamic relation $\;sT=p+\rho\;$.  Therefore,  $s$ is as large as it could be in comparison  to $\rho$, implying  entropic dominance  \cite{Bmd} and correspondingly large quantum fluctuations.

How does this all connect  to the skeptic's description of the polymer model?
Recalling Eq.~(\ref{pr}) but now with positive pressure ($\;l_P=\sqrt{G}\;$ is
the Planck length),
\be
p_r \;= \;+\rho\;=\; \frac{1}{8\pi l_P^2 r^2 }\;,
\ee
and using the above reasoning to obtain the entropy density,
\be
s\;=\;\sqrt{\frac{2\pi \rho}{l_P^2}}\;=\;\frac{1}{2 r l_P^2}\;,
\ee
we find that  the mass and entropy  inside a sphere of radius $\;R<R_S\;$
are respectively
\be
m(r < R)\;=\; \frac{1}{2}\frac{R}{l_P^2}
\label{ElessR}
\ee
and
\be
S(r < R)\;=\; \pi \frac{R^2}{l_P^2}\;.
\label{NlessR}
\ee
These are indeed the  expected scaling relations  given that  each spherical slice
is supposed to  act like a  BH horizon (just like  for the skeptic). However, because these configurations lack  a reliable
description in terms of a (semi)classical metric,
the coordinate $r$ should be thought of as a fiducial coordinate rather than the  radial coordinate of a spherically symmetric, classical geometry.

It is worth noting some prominent  features of the polymer model
that are evident when the pressure is regarded  as  positive \cite{strungout,emerge}:  The outer boundary behaves classically like a horizon simply because the interaction strength of the strings scales with $\hbar$. However, when
Plank's constant is turned on,  then  bits of string will leak out with  an energy and at a rate that matches those of  Hawking particles.~\footnote{The string bits are subject to Hagedorn-scale excitations inside. However,
the average of the net bound energy per bit is that of a typical Hawking mode.}  In addition, this model has been shown to  saturate both the maximally allowed value of the Lyapunov exponent and the minimally allowed ratio of shear viscosity to entropy density, as  first conjectured in  \cite{MaldShenk} and \cite{KSS} respectively. Most of this analysis would be of  little interest to  a skeptic or a modernist; both of whom would insist on a well-defined metric for the interior.

The radial pressure is, of course,  a  significant difference between the
skeptic and postmodernist version of the polymer model: It has gone
from maximally negative   $\;p_r= -\rho\;$ in the former case to maximally
 positive  $\;p_r=+\rho\;$ in the latter. Either way, the transverse pressure
$p_{\perp}$ is vanishing on average. Our viewpoint is that
the latter version  provides the  more fundamental description;
however, the geometry can {\em not} be  sourced by an energy--momentum tensor with $\;p_r=+\rho\;$ because there is formally no room for entropy in the Einstein equations. This is yet another manifestation of the inadequacy of  classical geometry
(or even  semiclassical geometry for that matter) when it is  used  to describe  a  maximally entropic configuration.

The deviation between the two pressures, even if maximally large, is somewhat superficial; the real difference is one of perspectives.
This point has already been made in \cite{bookdill}, but we  are now well
positioned to sharpen this distinction:
The skeptic accepts that the interior
could contain matter  but is  otherwise agnostic about its nature. He is simply  describing  the interior in  terms of   pressure $p$ and energy density $\rho$ and insists that these be chosen in a way that evades the Buchdahl bound.
He soon finds out that  the only viable loophole   is a  maximally negative pressure. And, although this implies that the entropy is vanishing, he neglects this ``minor detail'' as his  interest is classical gravity and not quantum thermodynamics. Moreover, given that there is no spontaneous emission of radiation in
the skeptic's version, a vanishing entropy seems reasonable. In contrast, the postmodernist quite willingly accepts the exotic stringy description of the interior and could use it to reproduce the above analysis, including the positive pressure.  The sign of the pressure, however, is not a problem because for this
observer, lacking an adequate geometry,  the Buchdahl bound is not relevant.  This divergence in perspectives  would be similar to a scenario in which the accelerated expansion of the Universe  is really due to undetectable entropic matter and not a cosmological constant \cite{newdS}.

\begin{figure}[htb]
\centerline{\hspace{-6cm}\includegraphics[width=13.5cm]{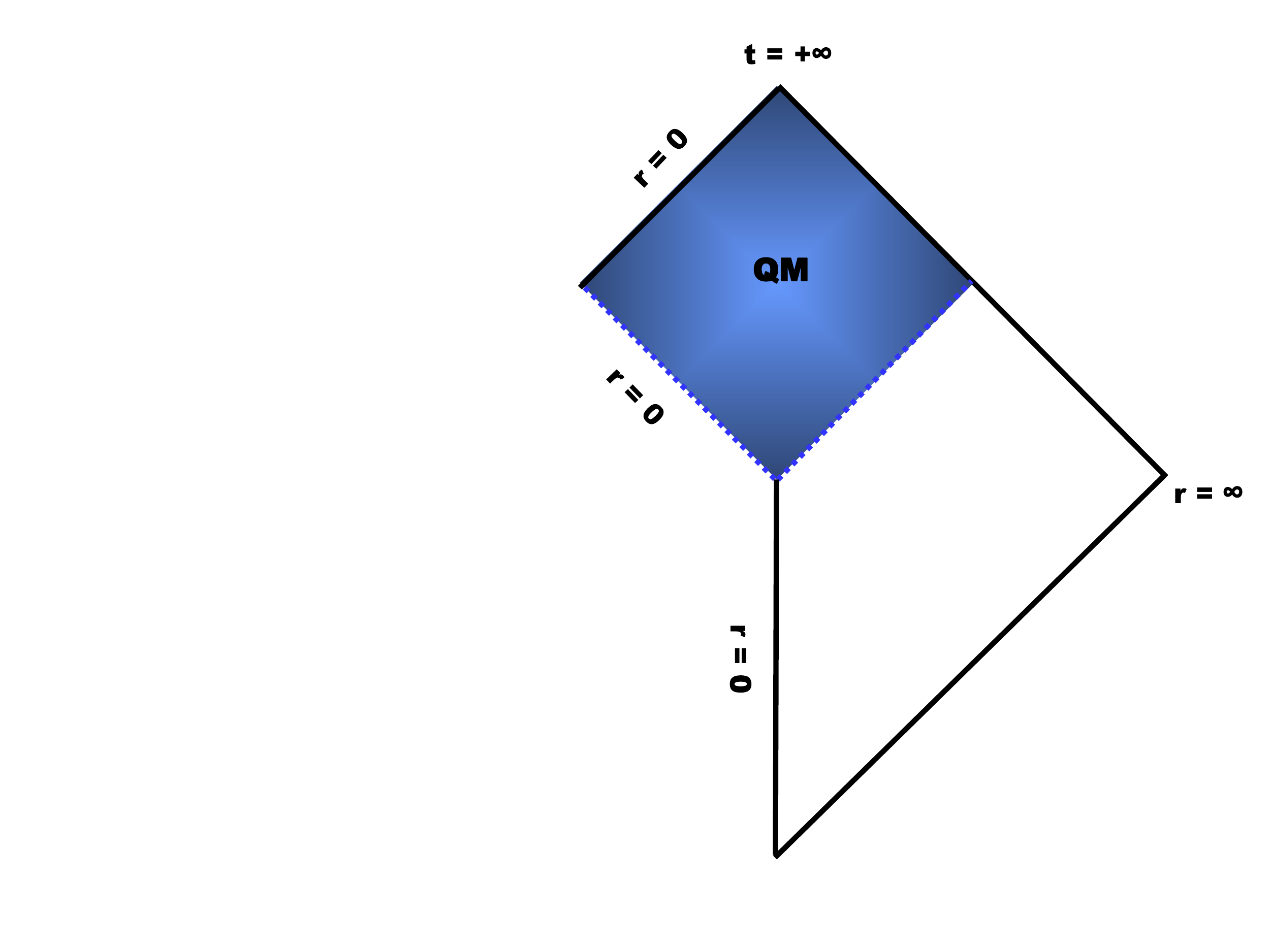}}
\caption{Penrose diagram of the polymer model of a regular BH from a postmodern  perspective.
See Fig.~\ref{fig:polymerskeptic} for the diagram representing the skeptic's  perspective of the same model. The above  diagram should also be viewed as part of an extended diagram.}
\label{fig:polymerPM}
\end{figure}

Although the two versions of the polymer model  are   different in some ways, they are quite  similar with regard to the propagation of signals in the interior. In both cases, all waves propagate at the speed of light because the magnitude of the pressure is maximal, $|p/\rho|=v^2/c^2$. One can also understand this from  physically motivated perspectives. In the skeptic's picture of the polymer model, the entire interior has to be  null and, therefore, so too must every  geodesic. Meanwhile, the postmodernist views the interior as being filled with long,  closed strings or, to some reasonable approximation, a single long loop of  string. As this is   effectively a $1+1$-dimensional entity, it follows that  all geodesics are,locally, maximally focused and any propagation must then follow the same path as a null ray.

The postmodernists  still need to explain why the polymer does not collapse, as  the negative-pressure argument no longer applies. These observers rather put an emphasis on the exotic nature of the interior matter  and discover that some simple calculations reveal a classically stable, Schwarzschild-sized state \cite{strungout}. They attribute this stability  to  the fact that the entropically favorable state for the strings is one of  a few  long loops  occupying a region of space  which  is the same size  as their  random walks. To see this, consider that
a long string of length $L$ has an entropy of $\;S_{long}=L/l_s\;$ (and a random-walk
size of $\sqrt{S_{long}}$),  whereas  $N$ shorter strings each of length $L/N$ have a total  entropy  of $\;S_{short}=\frac{\ln{N}}{N} S_{long}\;$, which is much smaller than $L/l_s$ even when $N$ is somewhat larger than one. For example, consider the case that the total entropy is that of a solar-mass BH,
 $\;S\sim 10^{76}\;$.  Then the difference between the entropy of a single long string and that of two half-length strings is $\frac{\ln{2}}{2}S$, which is of order of $S$ itself.  Moreover, this maximally large entropy explains  why an arbitrary state of matter
would transition into a hot stringy  soup in the first place.

How the polymer  evades the causality paradox  --- from the perspective of an external observer --- proved to be  the toughest nut to crack. Of course, an observer who accepts the complete picture, including the interaction rate of the strings, might not be so concerned. But can this issue be be reconciled  for the less informed as well? Fortunately,  this matter was recently resolved in \cite{ridethewave} as will now be explained.

We will start  with BHs out of equilibrium, paralleling part of  the discussion about the skeptic's perspective from the previous section. It  will be  assumed that the BH has been  excited by some external perturbation with  a monopole component that is either negligibly small  or  negative; both cases  are physically acceptable as observed  in \cite{ridethewave}. Such disturbances   will depress some part of the horizon via tidal effects \cite{Hartle, OSH,YotamNew}; in which case, an interior mode could very well  be exposed to the outside and thus have the opportunity to escape.   From an external perspective, a mode of  Hawking radiation could just as well have been  ``born'' in the exterior spacetime where causality is  certainly not an issue.

But what about a BH in equilibrium which is not exposed to any deforming sources; should it not be permitted to radiate? Recall that, from the skeptic's perspective, such a BH does not radiate. But the postmodernists want to include Hawking radiation in the picture, which means explaining  how a BH in equilibrium
can  overcome the causality paradox. For this, they  can rely on  quantum fluctuations in the horizon position, as the outer boundary of an ultracompact object would  normally experience Planckian-sized fluctuations at the very least \cite{emerge}. Probing a  sub-Planckian length scale would
turn the probing apparatus into a BH, so that one could never
know any position more precisely than $l_P$.
And, as explained  in \cite{ridethewave},
a Planckian-sized fluctuation is exactly what is needed. This is because a mode
of wavelength $l_P$ will be redshifted by a factor of $R_{S}/l_P$ by the time
it reaches an asymptotic observer.

The previous argument would seem to apply to just about any model of the interior. But, in actuality, one needs a fluid-filled object to account for the perpetual supply of near-horizon modes or, failing that, at least a limited concentration of matter in the center. This argument is then negated as far as a modernist is concerned. Meanwhile, it is the lack of quantum fluctuations in  the interior that makes it problematic for a skeptic to use. This is certainly the case
for the negative-pressure version of the polymer (which is completely resistant  to such fluctuations),
but also in general insofar as
regularized  BHs   are rendered  inconsistent by quantum effects.

\subsubsection*{The minimalist perspective}

A subgroup of the postmodernists  are those who would do away with the interior altogether, a school of thought which is akin to minimalism.  This would include Mathur's fuzzball model, according to his interpretation of the physical
picture   in
\cite{MathurNI} and \cite{MathurNI2}.  More emphatically, 't~Hooft argues away the interior by asserting that  incoming particles can be exchanged for outgoing ones before the energies of the former are red-shifted to Planckian scales; a process which he refers to as ``firewall transformations'' \cite{HooftX,HooftXX}. 't~Hooft also argues on behalf of an antipodal identification between the two exterior regions in the extended Penrose diagram. In this way, an incoming particle is  transmuted at the bifurcation surface and emerges in the opposite region without traversing
the interior. The two interior regions are  thus rendered irrelevant and can  be discarded. The reason that this model is classified under  postmodernism is because
both critical features are motivated by the requirement of preserving  quantum-mechanical principles. It will be argued elsewhere that
this description of the BH and that of the polymer model share much in common \cite{auntiewho}, which can be viewed as Hawking's principle of ignorance at work.

\subsection*{Conclusion}

Given Hawking's principle of ignorance and the
opaque nature of a BH horizon, different  observers do not necessarily  have to agree on what lies within.  With
this idea in mind, we have  considered the perspectives of three different external observers: the modernist, who expects something akin to the BHs of GR, the skeptic, who would prefer to eliminate  gravitational collapse but without giving up on  GR and  the postmodernist, who is willing to rely on exotic physics to achieve a self-consistent  quantum-mechanical description which is devoid of collapse. The preferences of each of the three observers are compared in Table~1.

\begin{table}[tbh]
\begin{center}
    \begin{tabular}{| l | c | c | c |c|c|}
    \hline & & & \\[-1.9em]
    \text{Observer} & \text{Regular} &\ \  \text{TD} \ \ &\ \ \text{QM}\ \  &\ \ \text{GR} \ \  & \ \ \text{SM}\ \  \\ [0.4em] \hline & & & \\[-1.9em]
    \text{Modernist} & No & Yes & No & Yes & Yes \\ [0.8em]
    \text{Skeptic} & Yes & No & No & Yes & No \\ [0.8em]
    \text{Postmodernist} & Yes & Yes  & Yes & No & No \\ [0.4em]
    \hline
    \end{tabular}
    \caption{Comparison between the three observers. The new  abbreviations TD and QM
stand for thermodynamics and  quantum mechanics, respectively.}\label{tab:1}
    \end{center}
    \end{table}

But, out of the three,  only the postmodern perspective  can consistently explain what
is known (so far) about BH physics. And, although the
discussion  has been emphasizing  the issues of gravitational collapse, reproducing BH thermodynamics and the causality paradox, any quantum-based, non-collapsing model of the interior is guaranteed to satisfy the quantum laws of unitary evolution and the strong subadditivity of entropy. Meaning that the notorious firewall problem is
similarly of no concern  to a  postmodernist. The singularity is resolved by
extending the quantum-gravitational scale all the way up to the horizon.

Taking the  principle of ignorance and the one-way nature
of the horizon seriously, one  might wonder if anything could be said definitively
about the BH interior. Fortunately, the ever-growing body of
gravitational-wave data casts a whole new light on this prospect. After all,
Hawking's principle stipulates that one's explanation must be consistent with what is already known, and so each new data point further constrains the collection of permissible stories. The reason that gravitational waves are particularly valuable in this way is because they represent the product of out-of-equilibrium physics  ---
for instance, the violent merger of the two BHs in a binary system ---  whereas the more
traditional aspects of BH physics (such as  Hawking radiation and the area--entropy law) assumes systems at or near equilibrium.  It is our  expectation
that any new physics should indeed leave just such a signature in the
gravitational-wave data \cite{collision}. Also  see \cite{Cardosorev,Carballo-Rubio:2018jzw} for a more  general  discussion.

One of the initial motivations for
this paper was to better understand how the inside
of a polymer BH can have two different equations of state
\cite{bookdill},
even though any relevant observer  would have to be located in the exterior.
But, as we now understand, the BH entropy may or may not be hidden depending on how the observer explains the
stability of the polymer against gravitational collapse.
It is our expectation that a similar dichotomy applies
to the accelerated expansion of the Universe \cite{newdS}.

Although a  main part of our focus was on the polymer BH, it should be
noted  that  models of BH-like objects are rampant in the literature (see \cite{Cardosorev} for a thorough yet incomplete catalog). But the takeaway point should be that exotic physics, by which we mean beyond the SM and GR, is necessary  for a consistent
description  of evaporating BHs.  Either way, the polymer BH would appear to be a viable model. But there will come a time when such speculations no longer  matter, as it does seem inevitable that any  given proposal will be seriously tested by data form future gravitational-wave experiments. It is then just as inevitable that, for the vast majority of BH models, the current state of ignorance will indeed  be  bliss.

\section*{Acknowledgments}

The research of RB   was supported by the Israel Science Foundation grant no. 1294/16. The research of AJMM received support from
an NRF Incentive Funding Grant 85353 and a
 Rhodes University
Discretionary Grant RD51/2018.
AJMM thanks Ben Gurion University for their  hospitality during his visit.



\end{document}